\documentclass{elsart}

\usepackage{amssymb}
\usepackage{amsmath}
\usepackage{amscd}

\newcommand{\eg}{\textit{e.g.}}
\newcommand{\ie}{\textit{i.e.}}

\begin{document}

\begin{frontmatter}

\title{The variant of ADHMN construction associated with $q$-analysis}

\author{Atsushi Nakamula}
\ead{nakamula@sci.kitasato-u.ac.jp}
\address{Department of Physics, School of Science, Kitasato University, Sagamihara 228-8555, Japan}

\begin{abstract}
A $q$-deformation of the ADHMN caloron construction is considered, under which  the anti-selfdual (ASD) conditions
 of the gauge fields are preserved.
It is shown that the $q$-dependent Nahm data with certain constraints are crucial to determine the ASD gauge fields, as in the 
case of ordinary caloron construction. 
As an application of the $q$-deformed ADHMN construction, we give a $q$-deformed caloron of Harrington-Shepard type.
Some limits of the parameters are also considered.
\end{abstract}

\end{frontmatter}

\section{Introduction}
The ADHM construction \cite{ADHM} is a vital tool to find out the exact solutions to the (anti-)selfdual (ASD) Yang-Mills equations in $\mathbb{R}^4$, explicitly. 
The construction is also effective for the instanton calculus of supersymmetric Yang-Mills theories, \eg \cite{KMS}.

On the compactified flat space  $\mathbb{R}^3\times S^1$, there exist 
the solutions  to ASD equations with a periodicity in the $S^1$ direction, \ie, the calorons, 
which have been discussed firstly in finite temperature field theories \cite{HS}.
Nahm \cite{Nahm} has applied ADHM's approach to the construction of calorons successfully
 by introducing infinite dimensional functional ($\mathcal{L}^2$) space as a dual space, the Nahm transformation. 
In the limit that  the size of constituent instanton  is sufficiently large compared with the circumference of the $S^1$, 
one can reproduce the monopole solutions to the Bogomolnyi equations \cite{Bog}, the ASD equations in $\mathbb{R}^3$.
On the other hand, the large circumference limit gives ordinary instanton solutions in $\mathbb{R}^4$.  
This aspect is strongly supported by the moduli space analysis of calorons \cite{Kr}.
We thus have an interpretation that calorons give the interpolation between instantons and monopoles \cite{ward1}.
There is another perspective of calorons that they may be interpreted as monopoles with a loop group as their gauge group \cite{GM,Norb}.

Some years ago, there have been found very interesting types of the calorons through
 the ADHM/Nahm (ADHMN) constructions.
One class of those is the calorons with non-trivial holonomy around $S^1$ at the spatial infinity \cite{LY,LL,KvB,BvB}, 
which brings gauge symmetry breaking through the Wilson loop mechanism.
The other class is the symmetric calorons \cite{ward1}, \ie, the multi-caloron solutions with certain spatial symmetries.
We expect that there is room for advanced applications of the ADHMN construction to discover new types of ASD solutions.    

In a series of papers \cite{KaNa}, Kamata and the present author have considered a $q$-analog of the BPS monopole construction, which gives an exact solutions to the ASD equation. 
They have used a variant formulation of the Nahm construction by introducing an $\ell^2$ functional space as a dual space instead of Nahm's $\mathcal{L}^2$ space, and 
obtained the solution with a parameter $q$ interpolating the BPS monopole $(q\to1)$ and the pure gauge configuration $(q\to0)$
, a $q$-deformed BPS monopole.
They also found that the  $q$-deformed BPS monopole could be interpreted as a special case of  the instantons with axial symmetry \cite{Wit}.

In this paper, we consider the generalization of the previous work  by formulating
the $q$-deformation of the ADHMN caloron construction, which contains the matrix Nahm data depending on $q$. 
As a concrete example, we will fix explicitly the vector $\vec{V}_q$ critical to give
 the $q$-deformed caloron of Harrington-Shepard type \cite{HS},
 which is an exact solution to the ASD equation on $\mathbb{R}^3\times S^1$.

This paper is organized as follows.
In section 2, we give the $q$-deformed ADHMN formalism which yields ASD gauge fields, following a brief introduction of the ADHMN caloron construction.
In section 3, we apply the formalism to yield a $q$-deformation of the Harrington-Shepard caloron.
Some limits of the parameters for the $q$-deformed caloron are also considered.
In the final section, we give concluding remarks and discussion.  

\section{The ADHMN construction on an $\ell^2$ functional space}
We consider the ADHMN construction in $Sp(1)$ formulation, \ie, the gauge group is restricted to $SU(2)$.
In Nahm's caloron construction, the $N\times N$ Nahm data are crucial to determine the ASD configuration,
which data are constrained by the following Nahm equations ($i=1,2,3$) with respect to the continous variable $z$,
\begin{equation}
\frac{dT_i}{dz}-\frac{1}{2}\sum_{j,k=1}^3\epsilon_{ijk}\left[T_j\,,T_k\right]-\left[T_i\,,T_0\right]=\frac{1}{2}\mbox{tr}_2\left(\sigma_i \Lambda^\dagger \Lambda\right)\delta(z-z_0),
\label{caloronNahm}
\end{equation}
 in addition to the anti-hermite conditions $T_i^\dagger=-T_i$.
In the right hand side, the trace is over quaternion and $\Lambda$ is an $N$ component quaternion valued row vector.
The distinction from the monopole Nahm data is that they are periodic in $z$ with period $2z_0:=2\pi/\beta$, 
where $\beta$ is a circumference of $S^1$.
The defining relations (\ref{caloronNahm}) are derived from the ASD conditions of the 1-dimensional ``Dirac operator",
\begin{equation}
\Delta=\left[\begin{array}{c}
\Lambda \delta(z-z_0) \\
i\partial_z+x+i\sum_{\mu=0}^3 T_\mu(z)\tau_\mu
\end{array}\right],
\end{equation}
\ie, $\Delta^\dagger\Delta$ is commutative with arbitrary quaternion and has inverse.
Here $x_\mu=(x_0,x_1,x_2,x_3)$ is a coordinate of $\mathbb{R}^3\times S^1$, $x_0$ being that of $S^1$,  $\tau_\mu=(1,i\sigma_1,i\sigma_2,i\sigma_3)$ is a quaternion element and $x:=\sum_{\mu=0}^3 x_\mu\tau_\mu$.
If the Nahm data are given, we determine a vector $\vec{V}$ of the form
\begin{equation}
\vec{V}=\left[\begin{array}{c}
V(x_\mu) \\
v(z,x_\mu)
\end{array}\right],\label{vector}
\end{equation}
where $V(x_\mu)$ is a quaternion valued function, and $v(z,x_\mu)$ is an $N$ component quaternion valued column vector periodic in $z$ with period $2z_0$,
that is, $v(z,x_\mu)$ being an element of  $\mathcal{L}^2[I]\otimes V_N$, where $I:=[-z_0,\,z_0]$.   
This vector is assumed to solve the differential equation taking into account the discontinuity at the boundary,
\begin{equation}
\Delta^\dagger\vec{V}=0\label{Diraceqn},
\end{equation}
or writing down (\ref{Diraceqn}) definitely,
\begin{equation}
\Lambda^\dagger\delta(z-z_0) V+(i\partial_z+x^\dagger+i\sum_{\mu=0}^3T_\mu\tau^\dagger_\mu)v(z)=0.\label{Diraceqn1}
\end{equation}
The procedure to find out the solution to (\ref{Diraceqn1}) is as follows: 
we solve the differential equation at $z\neq z_0$ to fix $v(z)$ firstly, then determine 
the top component $V$ by performing a short range integration at the boundary.  
In addition to (\ref{Diraceqn}), we impose a normalization condition  $\langle \vec{V},\vec{V}\rangle=1$,  
where the inner product  of the vector space with another vector $\vec{U}={}^t(U, u(z))$ is defined as, 
\begin{equation}
\langle \vec{U}, \vec{V}\rangle=U^\dagger V+\int_{-z_0}^{z_0}u^\dagger(z) v(z)dz.\label{L^2 inner prod}
\end{equation}
The gauge connection obeying ASD conditions can be determined by this normalized vector up to gauge transformation as,
\begin{equation}
A=\langle \vec{V}, d\vec{V}\rangle.\label{connection}
\end{equation}


We now consider a one parameter, $q$-, deformation of the ADHMN caloron construction
 which preserves the ASD condition of the gauge field.
In \cite{KaNa}, we have made such a deformation of the BPS monopole by introducing an $\ell^2$ functional space
 in place of $\mathcal{L}^2$ functional space.
We  apply the same approach to the caloron construction described above.
The differences to the previous work are the introduction of the $N\times N$ periodic Nahm data and the periodicity on $\vec{V}$.

We introduce an $\ell^2$ integrable periodic function $v(z;q)$ depending on a parameter $q\in(0,1)$ in place of 
the $\mathcal{L}^2$ integrable function considered in  Nahm's caloron construction.
The period is set to be $2z_0$, which is identical to the non-deformed case.
Here we define the $\ell^2$ integrability of $v(z;q)$ by the square integrability
 with the inner product defined on the infinite number of point set $I_q:=\{z_n,\, -z_n|\> z_n=q^nz_0,\, n=0,1,2,\dots\}$.
In the following, we will use the notation $I_q^{(\pm)}:=\{\pm z_n\}$, namely, $I_q=I_q^{(+)}\oplus I_q^{(-)}$.
We define the $\ell^2$ inner product of $u(z;q)$ and $v(z;q)$ as,
\begin{equation}
( u, v)_q=z_0(1-q)\sum_{n=0}^\infty\left(u^\dagger(z_{n+1};1/q) v(z_n;q) +u^\dagger(-z_{n+1};1/q) v(-z_n;q)\right)q^n,
\end{equation}
or by using the symbolic notation of the $q$-integration and the conjugate vector by $*$, the description can be simplified into
\begin{equation}
( u, v)_q=\int^{z_0}_{-z_0}u^*(qz) v(z) \,d_qz.\label{ell2}
\end{equation}
Notice that the shift of argument in the conjugate, or left, vector.
We, therefore, find that an $\ell^2[I_q]$ function $v(z;q)$ can be viewed as an infinite dimensional quaternion valued 
vector  $v(z;q)=v_+(z;q)\oplus v_-(z;q)$ with components
\begin{equation}
\begin{array}{c}
v_+(z;q)={}^t(v(z_0;q), v(z_1;q), v(z_2;q),\cdots),\\
v_-(z;q)={}^t(v(-z_0;q), v(-z_1;q), v(-z_2;q),\cdots) 
\end{array}
\end{equation}
and the $*$ conjugation of this quaternion valued vector $v^*(z;q)=v^*_+(z;q)\oplus v^*_-(z;q)$ being defined by
the hermite conjugation in addition to the ``twist" of $q$, 
\begin{equation}
\begin{array}{c}
v^*_+(z;q)=(v^\dagger(z_0;1/q), v^\dagger(z_1;1/q), v^\dagger(z_2;1/q),\cdots)\\
v^*_-(z;q)=(v^\dagger(-z_0;1/q), v^\dagger(-z_1;1/q), v^\dagger(-z_2;1/q),\cdots).
\end{array}\label{vstar}
\end{equation} 
Under this definition of the $\ell^2$ inner product, we can confirm the ``hermiticity" of $q$-difference operator $iD_z$ \cite{KaNa},
\ie, $(u,iD_zv)_q=(iD_zu,v)_q$, where $D_z\phi(z):=\phi(z)-\phi(qz)/z-qz$ for a function $\phi(z)$.

By using the $\ell^2[I_q]$ vector space introduced above, we now make a reformulation of  the ADHMN construction of calorons.
We define the ``Dirac operator" by applying the $q$-difference operator,
\begin{equation}
\Delta=\left[\begin{array}{c}
\Lambda \delta_q(z_0,z) \\
iD_z+x+i\sum_{\mu=0}^3 T_\mu(z;q)\tau_\mu
\end{array}\right],\label{q-delta}
\end{equation}
where $\delta_q(z,z_0)$ is given by a Kronecker delta rather than a delta function, 
\begin{equation}
\delta_q(z,z_0):=\frac{2}{(1-q)z}\delta_{z,z_0}.
\end{equation}
In analogy with the vector (\ref{vector}), the deformed vector is set to be of the form  
\begin{equation}
\vec{V}_q=\left[\begin{array}{c}
V(x_\mu;q) \\
v(z,x_\mu;q)
\end{array}\right],\label{q-vector}
\end{equation}
where $V(x_\mu;q)$ is a quaternion and $v(z,x_\mu;q)$ is a quaternion valued vector
 of $\ell^2[I_q]\otimes V_N$ rather than $\mathcal{L}^2[I]\otimes V_N$, and its conjugation is 
\begin{equation}
\vec{V}_q^*=
\left[
V^\dagger(x_\mu;1/q), \
v^*(z,x_\mu;q)
\right],
\end{equation}
the second component being defined by (\ref{vstar}). 
We define the inner product of those vectors by using (\ref{ell2}) as,
\begin{equation}
\langle \vec{U},\vec{V}\rangle_q:=U^\dagger(x_\mu;1/q) V(x_\mu;q)+(u,v)_q.
\end{equation}

In accordance with the ADHMN construction, the ASD gauge fields  are given by  
the condition that $\Delta^*\Delta$ is invertible and commutative with quaternion.
The explicit form of $\Delta^*\Delta$ is  given in the appendix. 
Here, we show the necessary conditions, the commutativity with quaternion, on the Nahm data
\begin{eqnarray}
\label{pseudo-constant} &T_i(z_n)=T_i(z_{n+1}),\ T_i(-z_n)=T_i(-z_{n+1}),&\\ 
&T_i^*(z_n)=-T_i(z_{n}),\ T_i^*(-z_n)=-T_i(-z_{n}),& \label{twist}\\
\label{q-Nahm} &D_zT_i-\frac{1}{2}\sum_{j,k=1}^3\epsilon_{ijk}\left[T_j\,,T_k\right]
-\left[T_i\,,T_0\right]
=\frac{1}{2}\mbox{tr}_2\left(\sigma_i \Lambda^\dagger \Lambda\right)\delta_q(z,z_0),&
\end{eqnarray}
where $i=1,2,3$.
By the conditions (\ref{pseudo-constant}), we find the Nahm data are ``pseudo-constant" in each interval $I_q^{(\pm)}$, \ie, 
$T_i(z_n)=T_i(z_0),\>T_i(-z_n)=T_i(-z_0)\> (n=1,2,\cdots)$ so we have omitted the argument in (\ref{q-Nahm}).
In spite of this, we should reserve
the $q$-difference term in (\ref{q-Nahm}) to pick up the contribution of the boundary discontinuity, 
which can exist since $0\notin I_q$.
Obviously, the pseudo-constant conditions do not imply that the $T_i$'s are strictly constant matrices in the continuous interval $I$.
The conditions (\ref{twist}) together with (\ref{pseudo-constant}) are ``twisted" anti-hermite conditions,
\begin{equation}
T^\dagger_i(z_0;q)=-T_i(z_0;q^{-1}),\ T^\dagger_i(-z_0;q)=-T_i(-z_0;q^{-1}).\label{twistedAH}
\end{equation} 
We find that (\ref{q-Nahm}) becomes the equations similar to the ADHM equation if $z\neq z_0$,
\begin{eqnarray}
\begin{array}{c}
\left[T_1,T_2\right]+\left[T_3,T_0\right]=0\>\\
\left[T_2,T_3\right]+\left[T_1,T_0\right]=0\>\\
\left[T_3,T_1\right]+\left[T_2,T_0\right]=0,
\end{array}\label{T0123}
\end{eqnarray}
where $T_i=T_i(\pm z_0)$ are pseudo-constant matrices on each $I_q^{(\pm)}$, respectively, whereas $T_0=T_0(z)$ is not constrained.
At present, we have no proof that $\Delta^*\Delta$ is invertible for the general Nahm data subject to the constraints
(\ref{pseudo-constant}), (\ref{twist}) and  (\ref{q-Nahm}),  because of the curious $\ell^2$ inner product.
The presence of the inverse $f:=(\Delta^*\Delta)^{-1}$ must be confirmed case by case after the Nahm data were
 fixed by the other constraints.
For the (non-deformed) Nahm construction of $SU(2)$ monopoles, and also calorons \cite{ward1}, the Nahm data must enjoy the residue conditions which guarantee  $\dim_\mathbb{H}\ker(\Delta^\dagger)=1$ \cite{Hit}.
In our deformed construction, the corresponding constraints $\dim_\mathbb{H}\ker(\Delta^*)=1$ have to be confirmed as well.

As in the ordinary caloron construction, if the Nahm data are given, the vector $\vec{V}_q$ is determined by
the  $q$-difference equation and the normalization condition,
\begin{eqnarray}
\label{q-dirac}\Delta^*\vec{V}_q=\Lambda^*\delta_q(z_0,z) V+(iD_z+x^\dagger+iT_\mu\tau^\dagger_\mu)v(z)=0\\
\label{q-norm}\langle \vec{V}_q,\vec{V}_q\rangle_q=V^\dagger(q^{-1})V(q)+(v,v)_q=1.
\end{eqnarray}
The connection one-form is thus given by 
\begin{equation}
A=\langle \vec{V}_q,d\vec{V}_q\rangle_q.
\end{equation}
By construction, this reduces to (\ref{connection}) in the limit $q\to1$.
Finally, since we are considering $\Delta$  in the standard form,
 $\Delta=\Delta_0+{}^t[0,x]$ for $\Delta_0$ independent of $x$,   
the curvature two-form $F=dA+A\wedge A$ is given by the canonical way,
 with the inverse $f$ and the ASD t'Hooft tensor $\bar{\eta}_{\mu\nu}$ such that,
\begin{equation}
F=\int^{z_0}_{-z_0} d_qz\int^{z_0}_{-z_0} d_qw\, v^*(qz)\bar{\eta}_{\mu\nu}f(z, qw)v(w) \>dx^\mu\wedge dx^\nu \label{F}.
\end{equation}
We can regard (\ref{F}) that the ordinary matrix product in the original ADHM construction is replaced by the $q$-integral.

\section{Example: a $q$-deformed caloron}
In this section, we explicitly apply the $q$-deformed ADHMN caloron construction on $\ell^2$ vector space formulated in the previous section.
The Nahm data to be considered here are the simplest case, $N=1$, in which case the data can be chosen,  
\begin{equation}
T_\mu(z;q)=0, \ \Lambda=\lambda\cdot1_2\,\label{Nahm data}
\end{equation}
where $\lambda\in\mathbb{R}$ and $1_2$ is a real quaternion element.
The first ansatz shows no boundary discontinuity in the Nahm data, which leads to the second one. 
In this case $\dim_\mathbb{H}\ker(\Delta^*)=1$ and there exists $f=(\Delta^*\Delta)^{-1}$, whose exact form is
the same as the $q$-deformed monopole case \cite{KaNa}.
The ASD conditions for the gauge field constructed by the Nahm data (\ref{Nahm data}) are, therefore, fulfilled.
We will find that this construction gives the $q$-deformed version of Harrington-Shepard caloron.

Having obtained the Nahm data, we next solve the difference equation (\ref{q-dirac}) to find 
the unnormalized vector $\vec{V}^0_q={}^t(V^0,v^0(z;q))$,
\begin{equation}
\lambda\delta_q(z_0,z) V^0+(iD_z+x^\dagger)v^0(z;q)=0.\label{qHSdirac}
\end{equation}
If $z\neq z_0$, (\ref{qHSdirac}) is reduced to a linear homogeneous $q$-difference equation $(iD_z+x^\dagger)v(z)=0$ so that
 $v$ is easily solved by, 
\begin{equation}
v^0(z,x_\mu;q)=e_q(i(1-q)zx^\dagger)\label{v^0}
\end{equation} 
where a $q$-exponential function $e_q(w)$ convergent at $|w|<1$ is 
\begin{equation}
e_q(w):=\sum_{n=0}^\infty\frac{w^n}{(q;q)_n}=\frac{1}{(w;q)_\infty},\label{e_q}
\end{equation}
and the $q$-shifted factorial $(a;q)_n=\prod_{k=1}^{n}(1-aq^{k-1})$.
It can be shown \cite{GR} that the second equality in (\ref{e_q}) follows from the $q$-binomial theorem, 
and that $\lim_{q\to1-} e_q((1-q)w)=e^w$.
To find the first component, $V^0$, we implement the short $q$-integral ($\epsilon\to0$) around the boundary $z_0$ taking into account 
the periodicity of $v(z;q)$,
\begin{equation}
0=\int_{z_0-\epsilon}^{z_0+\epsilon}\left(\lambda\delta_q(z_0,z) V^0+(iD_z+x^\dagger)v^0(z;q)\right)d_qz
=\lambda V^0+i\int_{z_0-\epsilon}^{z_0+\epsilon}D_zv^0(z;q)d_qz.
\end{equation}
The second term of the right hand side can be evaluated by using the ``fundamental theorem of $q$-calculus", thus,
\begin{eqnarray}
\int_{z_0-\epsilon}^{z_0+\epsilon}D_zv^0\,d_qz=\int_{z_0-\epsilon}^{z_0}D_zv^0\,d_qz+\int_{-z_0}^{-z_0+\epsilon}D_zv^0\,d_qz\nonumber\\
=(v^0(z_0)-v^0(z_0-\epsilon))+(v^0(-z_0+\epsilon)-v^0(-z_0))\nonumber\\
=v^0(-z_0+\epsilon)-v^0(z_0-\epsilon),\label{Vpure}
\end{eqnarray}
where we have used the periodicity $v(z+2z_0)=v(z)$.
The boundary value of (\ref{v^0}) together with the $q$-binomial theorem \cite{GR} gives, by using 
the spacetime variable $\rho_\pm:=x_0\pm ir$ for $r^2=\sum_{i=1}^3x_i^2$ and a quaternion 
$\hat{\mathbf{x}}=\sum_{i=1}^3x_i\sigma_i/r$, that
\begin{equation}
V^0=-\frac{i}{\lambda}\left(R_+(\rho_+;q)(1-\hat{\mathbf{x}})+R_-(\rho_-;q)(1+\hat{\mathbf{x}})\right), 
\end{equation}
where 
\begin{equation}
R_\pm=\frac{1}{2}\left(\frac{1}{(i\rho_\pm(1-q)z_0;q)_\infty}-\frac{1}{(-i\rho_\pm(1-q)z_0;q)_\infty}\right),\\
\end{equation}
and whose conjugation is
\begin{equation}
R^*_\pm=\frac{1}{2}\left((i\rho_\mp(1-q)z_0;q)_\infty-(-i\rho_\mp(1-q)z_0;q)_\infty\right).\\
\end{equation}

So we find that the normalized vector $\vec{V}_q$ should be  of the form 
\begin{equation}
\vec{V}_q=\vec{V}^0_q\,\phi^{-1/2}=\left[\begin{array}{c}
-\frac{i}{\lambda}\left(R_+(\rho_+;q)(1-\hat{\mathbf{x}})+R_-(\rho_-;q)(1+\hat{\mathbf{x}})\right) \\
e_q(ix^\dagger(1-q)z)
\end{array}\right]\phi^{-1/2},\label{V_q}
\end{equation}
where $\phi^{-1/2}$ is a quaternion valued normalization function which will be fixed by (\ref{q-norm}), \ie, 
$\left(\phi^{-1/2}\right)^*\langle \vec{V}^0_q, \vec{V}^0_q\rangle_q\phi^{-1/2}=1$.
The procedure to fix $\phi^{-1/2}$ is in the similar manner to the $q$-deformation of BPS monopole case \cite{KaNa}.
We finally find,
\begin{eqnarray}
\phi^{-1/2}=\frac{1}{2}\left\{(K_+-K_-+\Lambda_+-\Lambda_-)^{-1/2}+(K_++K_-+\Lambda_++\Lambda_-)^{-1/2} \right\}1\nonumber\\ 
-\frac{1}{2}\left\{(K_+-K_-+\Lambda_+-\Lambda_-)^{-1/2}-(K_++K_-+\Lambda_++\Lambda_-)^{-1/2} \right\}\hat{\mathbf{x}},
\nonumber\\
\label{N}
\end{eqnarray}
where the functions $\Lambda_\pm$ and $K_\pm$ are 
\begin{eqnarray}
\label{Lpm}&\Lambda_\pm=\frac{1}{i\rho_+}\sum_{n=0}^\infty\frac{\left(q\frac{\rho_-}{\rho_+};q\right)_{2n}}{\left(q;q\right)_{2n+1}}
\left(i(1-q)\rho_+z_0\right)^{2n+1}\pm(\rho_+\leftrightarrow \rho_-)\\
&K_\pm=\frac{2}{\lambda^2}(R_+^*R_+\pm R_-^*R_-).\label{Kpm}
\end{eqnarray}
In this way, we have completed fixing the vector $\vec{V}_q$ by the ADHMN construction on $\ell^2$ vector space, 
which yields the $q$-deformation of Harrington-Shepard caloron as will be shown later.
The anti-selfduality of the gauge field is obvious from (\ref{F}) as in the $q$-deformed monopole case \cite{KaNa}.
The entire information on the gauge connection and the curvature is included in the vector $\vec{V}_q$, 
so that they can be obtained by the canonical way but are very complex form.
Hereafter we consider $\vec{V}_q$ at some limits of the parameters.

Firstly, it can be shown that the $q\to1$ limit gives the ordinary Harrington-Shepard caloron.
In this limit, the $q$-integral turns into the ordinary integral so that the $\ell^2$ function getting close to an $\mathcal{L}^2$ function.
Actually, since 
\begin{equation}
\lim_{q\to1}K_+=\frac{2}{\lambda^2}(\cosh2z_0r-\cos2z_0x_0),\quad \lim_{q\to1}K_-=0,
\end{equation}
and 
\begin{equation}
\lim_{q\to1}\Lambda_+=\frac{\sinh2z_0r}{r},\quad \lim_{q\to1}\Lambda_-=0,
\end{equation}
we find the normalization function tends to
\begin{equation}
\lim_{q\to1}\phi^{-1/2}=\phi^{-1/2}(\mbox{HS})=\left(\frac{2}{\lambda^2}(\cosh2z_0r-\cos2z_0x_0)+\frac{\sinh2z_0r}{r}\right)^{-1/2}.
\end{equation}
The vector $\vec{V}_q$, therefore, turns into
\begin{equation}
\lim_{q\to1}\vec{V}_q=\left[\begin{array}{c}
\frac{2}{\lambda}\left(\sin x_0z_0\cosh rz_0-i\cos x_0z_0\sinh rz_0 \hat{\mathbf{x}}\right) \\
e^{ix^\dagger z}
\end{array}\right]\phi^{-1/2}(\mbox{HS}),
\end{equation}
which gives the Harrington-Shepard caloron  exactly through the $\mathcal{L}^2$ inner product (\ref{L^2 inner prod}).
As pointed out in \cite{ward1}, this caloron yields the interpolation between the 1-instanton in $\mathbb{R}^4$ (as $z_0\to0$)
 and the BPS 1-monopole (as $\lambda\to\infty$).
On the other hand, the  $q$-deformed caloron at general $q$ does not have the instanton limit $(z_0\to0)$,
 since the $q$-interval does not include zero, $0\notin I_q$.

Next, we consider the limit $\lambda\to\infty$, corresponding to the infinite instanton size, which is easily obtained from (\ref{V_q}) and (\ref{N}),
\begin{equation}
\lim_{\lambda\to\infty}\vec{V}_q=\left[\begin{array}{c}
0 \\
e_q(ix^\dagger(1-q)z)
\end{array}\right]\phi^{-1/2}(q\mbox{-BPS}),
\end{equation}
where
\begin{eqnarray}
\phi^{-1/2}(q\mbox{-BPS})=\lim_{\lambda\to\infty}\phi^{-1/2}=
\frac{1}{2}\left\{(\Lambda_+-\Lambda_-)^{-1/2}+(\Lambda_++\Lambda_-)^{-1/2} \right\}1\nonumber\\
-\frac{1}{2}\left\{(\Lambda_+-\Lambda_-)^{-1/2}-(\Lambda_++\Lambda_-)^{-1/2} \right\}\hat{\mathbf{x}},
\end{eqnarray}
This is exactly the $q$-deformed BPS monopole \cite{KaNa}, which becomes the BPS 1-monopole when $q\to 1$,
 and the pure gauge  ($F=0$) when $q\to0$.

Finally, we show the $q\to0$ limit tends to the pure gauge, analogous to the ``$q$-BPS" case.
In our $q$-deformed caloron, the field strength two form is given by the double $q$-integral (\ref{F}).
In the $q\to0$ limit, this double integral is reduced to the summation only at the boundary ($z=\pm z_0$),
\begin{eqnarray}
&F\longrightarrow z_0^2\left(v^\dagger(z_1;1/q)\eta_{\mu\nu}f(z_0,z_1) v(z_0)\right.
+v^\dagger(-z_1;1/q)\eta_{\mu\nu}f(-z_0,z_1) v(z_0)\nonumber\\
&+v^\dagger(z_1;1/q)\eta_{\mu\nu}f(z_0,-z_1) v(-z_0)\nonumber\\
&\left.+v^\dagger(-z_1;1/q)\eta_{\mu\nu}f(-z_0,-z_1) v(-z_0)
\right) dx^\mu\wedge dx^\nu,
\end{eqnarray}
however, from the boundary condition of $f(z,w)$, this is automatically zero.
 
In summary, we have had the following diagram.
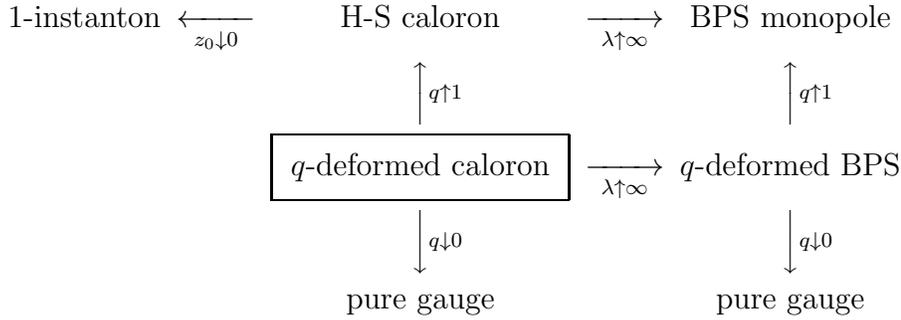
\[
\begin{CD}
\mbox{1-instanton} @<<{z_0\downarrow0}<\mbox{H-S caloron} @>> {\lambda\uparrow\infty}> \mbox{BPS monopole}\\
@. @AA{q\uparrow1}A                                                        @AA{q\uparrow1}A               \\
@. \mbox{\fbox{$q$-deformed caloron}} @>> {\lambda\uparrow\infty}> \mbox{$q$-deformed BPS} \\
@. @VV{q\downarrow0}V                                   @VV{q\downarrow0}V \\
@.\mbox{pure gauge}  @. \mbox{pure gauge}
\end{CD}
\]

\section{Concluding remarks}
In this paper, we have constructed the ADHMN caloron construction on $\ell^2$ vector space, 
which is a generalization of \cite{KaNa}. 
We have found the ASD necessary conditions on the matrix Nahm data, which are  (\ref{pseudo-constant}), (\ref{twist}) and (\ref{q-Nahm}).
As a concrete example, we have made the $q$-deformed Harrington-Shepard caloron, which preserves the ASD conditions for general
value of $q\in(0,1)$.

Further application of the $q$-deformed caloron construction will be possible.
It is intriguing to consider the $q$-deformed Nahm data for general matrix dimensions.
In contrast to the $N=1$ case considered in this paper, the $q$-deformed Nahm data of $N>1$ have  no counter part in the ordinary
caloron construction at $q\to1$, due to the pseudo constant condition (\ref{pseudo-constant}).
Such $q$-deformed Nahm data will be corresponding to a new class of gauge fields.
As a naive example of $N=2$ case, we can take an ansatz such as $T_0=T_1=T_2=0,$ and $T_3=iC(q)\sigma_3$,
 where $C(q)$ being a pseudo-constant with twist invariance, $C(q^{-1})=C(q)$,
and $\Lambda$ to be a real quaternion, 
which Nahm data gives the solution to (\ref{pseudo-constant}), (\ref{twist}) and  (\ref{q-Nahm}).
The inverse of $\Delta^*\Delta$ can be constructed as in the $N=1$ case, so that the gauge field is ASD.
However, the Nahm data do not have $q\to 1$ counter part obviously.
In fact, this leads to $\dim_\mathbb{H}\ker(\Delta^*)=2$, not to give an $Sp(1)\simeq SU(2)$ gauge field.
From this consideration, the generalization to higher rank gauge group must be considered.

Finally, it will be straightforward to construct a $q$-deformed calorons of $N=1$ with nontrivial holonomy at the spatial infinity by introducing the extra discontinuous points of the Nahm data  in the $q$-interval $I_q$, similarly to \cite{LL}.
This $q$-deformation will bring us the perspective of the $q$-caloron from the constituent $q$-monopoles, which is also quite
interesting subject to study.

\section*{Acknowledgement}
The author thanks Prof. Masaru Kamata for fruitful discussions.
He is also grateful to the participants of the workshop ``Fundamental Problems and Application of Quantum Field Theory" held 
at Yukawa Institute of Theoretical Physics in December 2005.
This work is partially supported by Grant-in-Aid for Scientific Research from Japan Society for the Promotion of Science No.16540352.

\appendix

\section{The ASD conditions}

In this appendix, we give the explicit form of  $\Delta^*\Delta$, which is necessary to be commutative with arbitrary quaternion.
By using the formula of $q$-difference,
\begin{equation}
D_z\cdot\phi(z)=D_z\phi(z)\cdot+\phi(qz)\cdot D_z,
\end{equation}
we carry out the calculation of $\Delta^*\Delta$,
\begin{eqnarray*}
&\Lambda^*\Lambda\delta_q(z_0,z)
+(iD_z\otimes1_N+1\otimes1_N x^\dagger-1\otimes i\sum_{\mu=1}^4 T_\mu^*(z)\tau_\mu^\dagger)\\
&\times (iD_z\otimes1_N+1\otimes1_N x^\dagger+1\otimes i\sum_{\mu=1}^4T_\mu(z)\tau_\mu)\\
&=\Lambda^*\Lambda\delta_q(z_0,z)-D_z^2\otimes1_N+iD_z\otimes1_N2x_0+1\otimes1_N|x|^2+1\otimes T^*_0(z)T_0(z)\\
&-1\otimes D_zT_0(z)-D_z\otimes \left(T_0(qz)- T^*_0(z)\right)\\
&-1\otimes \sum_{k=1}^3 D_zT_k(z)\tau_k+1\otimes\sum_{j,k=1}^3 T^*_j(z)T_k(z)\tau_j^\dagger\tau_k\\
&-D_z\otimes\sum_{k=1}^3\left( T_k(qz)\tau_k+T^*_k(z)\tau_k^\dagger\right)\\
&+1\otimes i\sum_{k=1}^3\left(T_k(z) x^\dagger\tau_k-T^*_k(z)\tau_k^\dagger x\right)\\
&+1\otimes i\left(T_0(z) x^\dagger- T^*_0(z) x\right)\\
&+1\otimes \sum_{k=1}^3\left(T^*_0(z)T_k(z)\tau_k+T_k^*(z)T_0(z)\tau_k^\dagger\right).
\end{eqnarray*}
For the final formula to be commutative with quaternion, we need all of the terms including $\tau_k$, $x$ or $x^\dagger$ vanish.
Therefore, we find that it is necessary to hold  the pseudo constant condition (\ref{pseudo-constant}), the twisted anti-hermite condition (\ref{twist}) and the $q$-analog of the caloron Nahm equations, 
\begin{eqnarray}
D_zT_i(z)-\sum_{j,k=1}^3\epsilon_{ijk}T_j(qz)T_k(z)-T_i(qz)T_0(z)+T_0(z)T_i(z)\nonumber\\
=\Lambda^*\Lambda\delta_q(z_0,z),
\end{eqnarray}
which, together with (\ref{pseudo-constant}), yields (\ref{q-Nahm}). 
Finally we obtain the operator on $\ell^2[I_q]\otimes V_N$ without the boundary term,
\begin{eqnarray}
&\Delta^*\Delta=-D_z^2\otimes1_N+2x_0iD_z\otimes1_N+1\otimes1_N|x|^2\nonumber\\
&-1\otimes \sum_{\mu=1}^4(T_\mu(z)-2ix_\mu)T_\mu(z)-1\otimes D_zT_0(z)-D_z\otimes(T_0(qz)+T_0(z))\nonumber\\
\end{eqnarray}
in which each entry of the $N\times N$ matrices is a real quaternion.

\end{document}